\documentclass[reprint,amsmath,amssymb,superscriptaddress,prb]{revtex4-2}

\usepackage{graphicx}
\usepackage{dcolumn}
\usepackage{bm}
\usepackage{color}
\usepackage{amsmath}
\usepackage{booktabs} 
\usepackage{array} 
\usepackage{makecell}
\usepackage{colortbl}
\usepackage{xcolor} 
\usepackage{subcaption} 

\begin{document}

\title{\textit{Ab initio} study of highly charged ion–induced Coulomb explosion imaging}

\author{Misa Viveiros}
\affiliation{Department of Physics and Astronomy, Vanderbilt University, Nashville, TN 37235, USA}

\author{Samuel S. Taylor}
\affiliation{Pritzker School of Molecular Engineering, University of Chicago, Chicago, IL 60637, USA}

\author{Cody Covington}
\affiliation{Department of Chemistry, Austin Peay State University,
Clarksville, TN 37044, USA}

\author{K\'alm\'an Varga}
\email{kalman.varga@vanderbilt.edu}
\affiliation{Department of Physics and Astronomy, Vanderbilt University, Nashville, TN 37235, USA}

\begin{abstract}
We present a theoretical investigation of ion-induced Coulomb explosion imaging (CEI) of pyridazine molecules driven by energetic C$^{5+}$ projectiles, using time-dependent density-functional theory (TDDFT) with Ehrenfest nuclear dynamics. By systematically varying the projectile's impact point and orientation relative to the molecular plane, we compare orthogonal and in-plane trajectories and quantify their effects on fragment momenta, electron-density response, and atom-resolved ionization. Newton plots and time-resolved density snapshots show that trajectories avoiding direct atomic collisions yield the most faithful structural reconstructions, whereas direct impacts impart large, highly localized momenta that distort the recovered geometry. Planar trajectories generate substantially greater ionization and broader momentum distributions than orthogonal ones due to deeper traversal through the molecular electron cloud. Quantitative analysis of electron removal at 10~fs confirms that projectile proximity and orientation strongly modulate both local and global ionization. These findings clarify how impact geometry governs the fidelity of ion-induced CEI structural recovery and help explain the variability and noise observed in experimental CEI measurements. More broadly, the results highlight both the strengths and the intrinsic limitations of ion-induced CEI and identify key considerations for interpreting experiments.
\end{abstract}

\maketitle

\section{Introduction}
Coulomb explosion imaging (CEI) is a powerful technique for reconstructing complex molecular structures~\cite{doi:10.1126/science.244.4903.426, annurev:/content/journals/10.1146/annurev-physchem-090419-053627, PhysRevResearch.4.013029, PhysRevA.91.053424, Boll2022}. It has been successfully applied to distinguish isomers~\cite{Ablikim2016, Pathak2020} and enantiomers~\cite{doi:10.1126/science.1246549, doi:10.1126/science.1240362}, and to classify protein structures~\cite{André2025, PhysRevLett.134.128403}. In CEI, a molecule is multiply ionized so that the positively charged fragments repel each other, and the measured fragment momenta are then used to reconstruct the original molecular geometry. The technique is especially valuable for capturing ultrafast nuclear motion on near-instantaneous timescales (10--50~fs)~\cite{Li2021, PhysRevA.83.013417}, such as hydrogen migration~\cite{PhysRevLett.99.258302, XU2009255, XU2010119}, isomerization~\cite{C0CP02333G, Guo2024}, roaming dynamics~\cite{10.1063/1.5041381, Légaré_2006, doi:10.1126/science.abc2960, Mishra2024}, and other structural transformations~\cite{Zhang2022, Howard2023, Unwin2023, PhysRevA.101.012707}. 

CEI is typically performed using strong electric fields, which have been demonstrated to accurately 
image complex molecules on ultrafast timescales~\cite{doi:10.1126/science.adk1833, YATSUHASHI201852}. 
Recently, however, Yuan \textit{et al.}~\cite{PhysRevLett.133.193002} showed that CEI of complex 
molecules can be achieved with comparable accuracy using an energetic ion beam. 

In their study, a 112.5~keV/u C$^{5+}$ beam was used to induce Coulomb explosion in 
pyridazine, pyrimidine, and pyrazine---three isomers of 
C\textsubscript{4}H\textsubscript{4}N\textsubscript{2}. Analysis of the momentum distributions and 
angular correlations enabled accurate identification and differentiation between the tested isomers. 

Coulomb explosion imaging induced by energetic ion beams and by
intense femtosecond laser pulses rely on fundamentally different
ionization mechanisms, leading to complementary advantages and
limitations. In energetic ion–beam CEI, fast highly charged
projectiles strip multiple electrons from a molecule in an interaction
lasting only a few attoseconds to tens of attoseconds, producing
extremely high charge states and this
ultrashort, nearly impulsive ionization means that the measured
fragment momenta provide a direct snapshot of the nuclear geometry,
largely unaffected by field-induced alignment or molecular
polarization. In contrast, laser-driven CEI uses strong-field
tunneling or over-the-barrier ionization during a femtosecond pulse,
where the electric field simultaneously aligns, distorts, and ionizes
the molecule, giving rise to strong orientation selectivity and
ionization pathways that depend sensitively on the instantaneous laser
field. While laser CEI offers exquisite control via polarization,
pulse shaping, and pump–probe timing—enabling time-resolved tracking
of ultrafast structural dynamics—it typically reaches lower charge
states and may blur the initial geometry due to multi-step ionization
during the pulse.

Despite the promising experimental advances \cite{
PhysRevLett.133.193002,PhysRevA.103.042813,PhysRevA.101.042706}, a corresponding 
computational or theoretical investigation has not yet been performed. 
Such a study is essential for understanding the underlying mechanisms 
of ion-induced fragmentation and ionization, particularly the dependence 
on projectile impact point and molecular orientation.

The typical computational framework employed in CEI studies is based on classical 
Coulomb explosion models, where atoms are treated as point charges with fixed charge 
states that interact via purely classical electrostatic forces~\cite{PhysRevLett.133.193002, PhysRevLett.132.123201, lam2024imagingcoupledvibrationalrotational, C7CP01379E}. While computationally efficient, this approach does not capture time-dependent ionization, the detailed interaction of the projectile ion with the electronic structure of the target, or quantum effects that have been shown to influence CEI outcomes~\cite{PhysRevA.111.033109, PhysRevA.109.052813}. To accurately model the dynamics underlying ion-induced CEI, a more advanced computational tool is required.

Real-space, real-time time-dependent density functional theory (TDDFT) provides such a tool. TDDFT naturally incorporates quantum mechanical electron dynamics, allows for time-dependent ionization, and treats nuclear motion self-consistently via Ehrenfest dynamics~\cite{PhysRevA.111.033109}. As a result, it captures both the gradual depletion of electron density and the concurrent buildup of positive charge during ionization. TDDFT has been successfully applied to study time-dependent ionization~\cite{Livshits2006}, molecular rearrangement~\cite{PhysRevA.111.013109}, and electronic excitation~\cite{zkwl-sj59}, and has been validated against experimental results in laser-driven Coulomb explosion~\cite{Mogyorosi2025, PhysRevA.86.043407, PhysRevA.91.023422}, ion--molecule interactions~\cite{PhysRevB.85.235435, viveiros2025lowenergyprotonimpactdynamics, 10.1063/5.0296066, PhysRevA.95.052701}, and CEI-based structural reconstruction~\cite{PhysRevA.111.033109}.

In this work, we use TDDFT to investigate the mechanisms governing ion-induced CEI under conditions matching those of the experiment by Yuan \textit{et al.}~\cite{PhysRevLett.133.193002}. We examine the predicted momentum images, ionization behavior, orientation dependence, and the extent of molecular damage or fragmentation. We consider a C$^{5+}$ projectile ion with a kinetic energy of 1.35~MeV (equivalent to 112.5~keV/u for $^{12}$C) directed toward pyridazine, chosen as a representative system because it was included in the experimental study. To balance computational feasibility with coverage of relevant collision geometries, we sample a set of representative impact points located at chemically relevant sites, such as atomic centers, bonds, and the molecular center. This strategy captures key phenomena---including impact-point and orientation dependence---while maintaining manageable computational cost.

\section{Computational Method}
\label{sec:computational-method}

The simulations were performed using TDDFT for modeling the electron dynamics on a 
real-space grid with real-time propagation \cite{Varga_Driscoll_2011a}, 
with the Kohn-Sham (KS) Hamiltonian of the following form
\begin{equation}
\begin{split}
\hat{H}_{\text{KS}}(t) = -\frac{\hbar^2}{2m} \nabla^2 + V_{\text{ion}}(\mathbf{r},t) + 
V_{\text{H}}[\rho](\mathbf{r},t) \\
+ V_{\text{XC}}[\rho](\mathbf{r},t) + V_{\text{P}}(\mathbf{r},t).
\end{split}
\label{eq:hamiltonian}
\end{equation}
Here, $\rho$ is the electron density, defined as the sum of the densities of all occupied orbitals:
\begin{equation}
\rho(\mathbf{r},t) = \sum_{k=1}^{N_\text{orbitals}} 2 |\psi_k(\mathbf{r},t)|^2,
\end{equation}
where the coefficient 2 accounts for the number of electrons in each orbital $\psi_k$. 

$V_{ion}$ in Eq.~\ref{eq:hamiltonian} is the external potential due to the ions, represented by employing norm-conserving pseudopotentials centered at each ion as given by Troullier and Martins~\cite{PhysRevB.43.1993}. $V_{H}$ is the Hartree potential, defined as
\begin{equation}
V_H(\mathbf{r}, t) = \int \frac{\rho(\mathbf{r}', t)}{|\mathbf{r} - \mathbf{r}'|} \, d\mathbf{r}',
\end{equation}
and accounts for the electrostatic Coulomb interactions between electrons. $V_{XC}$, is the exchange-correlation potential, which is approximated by the adiabatic local-density approximation (ALDA), obtained from a parameterization to a homogeneous electron gas by Perdew and Zunger~\cite{PhysRevB.23.5048}. The last term in Eq.~\ref{eq:hamiltonian}, $V_{\text{P}}$ accounts for the Coulomb potential of the projectile.

At the beginning of the TDDFT calculations, the ground state of the system is prepared by performing a density-functional theory (DFT) calculation. With these initial conditions in place, we then proceed to propagate the KS orbitals, $\psi_{k}(\mathbf{r},t)$ over time by using the time-dependent KS equation, given as 
\begin{equation}
i \hbar \frac{\partial \psi_k(\mathbf{r}, t)}{\partial t} = \hat{H}_{\text{KS}}(t) \psi_k(\mathbf{r}, t).
\label{eq:tdks}
\end{equation}
Eq.~\ref{eq:tdks} was solved using the following time propagator
\begin{equation}
\psi_k(\mathbf{r}, t + \delta t) = \exp\left(\frac{-i \delta t}{\hbar} \hat{H}_{\text{KS}}(t) \right) \psi_k(\mathbf{r}, t).
\end{equation}
This operator is approximated using a fourth-degree Taylor expansion, given as
\begin{equation}
\psi_k(\mathbf{r}, t + \delta t) \approx \sum_{n=0}^{4} \frac{1}{n!} \left(\frac{-i \delta t}{\hbar} \hat{H}_{\text{KS}}(t)\right)^n \psi_k(\mathbf{r}, t).
\end{equation}
The operator is applied for $N$ time steps until the final time, $t_{final} = N \cdot \delta t$, is obtained. In our simulations, a time step of $\delta t = 1$~attosecond (as) and a total propagation time of $t_{\text{final}} = 60$~femtoseconds (fs) were used.

In real-space TDDFT, the KS orbitals are represented at discrete points on a uniform rectangular grid in real space. The simulation accuracy is governed by the grid spacing. In our calculations, we employed a grid spacing of 0.3~\AA\ and used 100 grid points along each of the $x$-, $y$-, and $z$-axes.

To enforce boundary conditions, we set the KS orbitals to zero at the edges of the simulation cell. However, during ion impact events, the collision can impart sufficient energy to the electronic wavefunctions, potentially leading to ionization or the ejection of electronic density beyond the molecule. In such cases, unphysical reflections of the wavefunction from the cell boundaries can occur, introducing artifacts into the simulation. To mitigate this issue, we implemented a complex absorbing potential (CAP) to dampen the wavefunction as it approaches the boundaries. The specific form of the CAP used in our simulations, as described by Manolopoulos~\cite{10.1063/1.1517042}, is given by:
\begin{equation}
- i w(x) = -i \frac{\hbar^2}{2m} \left(\frac{2\pi}{\Delta x}\right)^2 f(y),
\end{equation}
where $x_{1}$ is the start and $x_{2}$ is the end of the absorbing region, $\Delta x = x_{2} - x_{1}$, $c = 2.62$ is a numerical constant, $m$ is the electron’s mass, and
\begin{equation}
f(y) = \frac{4}{c^2} \left( \frac{1}{(1 + y)^2} + \frac{1}{(1 - y)^2} - 2 \right), \quad y = \frac{x - x_1}{\Delta x}.
\end{equation}

As the molecule becomes ionized during the simulation, electron density is driven towards the CAP. Additionally, any ejected fragments carry their associated electron density as they move towards the boundaries. When electron density reaches the CAP region, it is absorbed. Consequently, the total number of electrons,
\begin{equation}
N(t) = \int_V \rho(\mathbf{r}, t) \, d^3x,
\end{equation}
where \(V\) is the volume of the simulation box, decreases relative to the initial electron number, \(N(0)\). In our simulations, the CAP was applied at -7.5~\AA\ to 7.5~\AA\ along each Cartesian axis, allowing 15~\AA\ of free electron dynamics within the central region before the absorbing potential at the grid boundaries takes effect.

To determine the number of electrons associated with a specific atom or molecular fragment, we integrate the time-dependent electron density over a finite spatial region surrounding that atom. This procedure naturally yields non-integer electron counts. Such non-integer values are a well-known consequence of using local and semi-local exchange--correlation (XC) approximations in TDDFT, such as ALDA. The exact XC potential contains spatial steps and derivative discontinuities that enforce integer electron localization during dissociation, but these features are absent in ALDA. As a result, approximate XC potentials allow fractional electron delocalization and fractional charge transfer, especially in strong-field or ionizing environments~\cite{PhysRevA.91.023422,PhysRevA.95.052701,PhysRevA.86.043407}.

These fractional electron numbers can be interpreted in several physically meaningful ways. For instance, if integrating the electron density around a C atom yields 3.40 valence electrons (corresponding to a charge of $+0.60$), one interpretation is that approximately 3.40 electrons remain localized within the atomic region during the simulation, with the fractional charge either recombining with the ionized electron cloud or dissociating over a longer simulation period. Alternatively, a more practical interpretation is statistically: the C fragment may appear as neutral ($4$ valence electrons) in some realizations and as C$^{+}$ ($3$ valence electrons) in others, with the measured value representing the average. In this view, a count of 3.40 electrons corresponds to a fragment that would be detected with a charge of $+1$ ion in approximately 60\% of events and as a neutral atom in the remaining 40\%.

Motion of the ions in the simulations were treated classically. Using the Ehrenfest theorem
, the quantum forces on the ions due to the electrons are given by the derivatives 
of the expectation value of the total electronic energy with respect to the ionic positions. 
These forces are then fed into Newton’s Second Law, giving
\begin{equation}
\begin{split}
M_i \frac{d^2 \mathbf{R}_i}{dt^2} = \sum_{j \neq i}^{N_{\text{ions}}} \frac{Z_i Z_j (\mathbf{R}_i - \mathbf{R}_j)}{|\mathbf{R}_i - \mathbf{R}_j|^3}\\ 
- \nabla_{\mathbf{R}_i} \int V_{\text{ion}}(\mathbf{r}, \mathbf{R}_i) \rho(\mathbf{r}, t) \, d\mathbf{r},
\end{split}
\label{eq:newton-law-ions}
\end{equation}
where $M_{i}$, $Z_{i}$, and $\mathbf{R}_{i}$ are the mass, 
pseudocharge (valence), and position of the $i^{th}$ ion, respectively, and $N_{\text{ions}}$ is the total number of ions. 
$V_{\text{ion}}(\mathbf{r},\mathbf{R}_i)$ is the pseudopotential representing the combined effect of the nucleus and core electrons, and it interacts with the electron density $\rho(\mathbf{r}, t)$ via Ehrenfest dynamics.

Since the projectile speed is much greater than the nuclei in the molecular system, the forces experienced by the projectile were represented as being strictly Coulombic, given as
\begin{equation}
\begin{split}
M_p \frac{d^2 \mathbf{R}_p}{dt^2} = \sum_{i \neq p}^{N_{\text{ions}}} \frac{Z_i Z_p (\mathbf{R}_i - \mathbf{R}_p)}{|\mathbf{R}_i - \mathbf{R}_p|^3}\\ 
- Z_{p} \int \rho(\mathbf{r}, t) \frac{(\mathbf{r}-\mathbf{R}_p+\beta)}{|\mathbf{r}-\mathbf{R}_p+\beta|^3} \, d\mathbf{r},
\end{split}
\label{eq:newton-law-sc}
\end{equation}
where $\mathbf{R}_p$ is the position of the projectile and $\beta$ is the soft Coulomb (SC) parameter described below. The equations of motion for the molecular nuclei (Eq.~\ref{eq:newton-law-ions}) and the projectile (Eq.~\ref{eq:newton-law-sc}) were propagated in time using the Verlet algorithm, with time steps of $\delta t = 1~\mathrm{as}$ for the nuclei and $\delta t = 0.005~\mathrm{as}$ for the projectile. The projectile requires 200 times more integration steps than the nuclei due to its significantly higher velocity, necessitating a finer time resolution for accurate propagation.

Because a real-space grid is employed, the Coulombic potential of the projectile, \(V_{\text{P}}\) in Eq.~\ref{eq:hamiltonian}, is represented using a soft Coulomb (SC) potential~\cite{PhysRevA.44.5997, PhysRevA.80.032507}:
\begin{equation}
V_{\text{P}}(\beta , Z; \mathbf{r}, t) = -\frac{Z}{\sqrt{|\mathbf{r} - \mathbf{R}_p(t)|^2+\beta^{2}}}.
\end{equation}
Here, $Z$ is the charge of the ion projectile, and $\beta$ is the softening parameter introduced to avoid the singularity at $\mathbf{r} = \mathbf{R}_p(t)$. Ideally, $\beta$ should be as small as the grid spacing permits, since larger values alter the system’s energy levels and ionization potential. In our simulations, we set $\beta$ to approximately the real-space grid spacing (0.3~\AA), balancing the need to avoid the Coulomb singularity with minimizing distortion of the potential~\cite{PhysRevA.95.052701,G_L_Ver_Steeg_2003}. This approach has been successfully applied in previous studies of ion-molecule interactions~\cite{PhysRevA.95.052701}.

\begin{figure}[ht!]
    \centering
    \includegraphics[width=0.95\columnwidth]{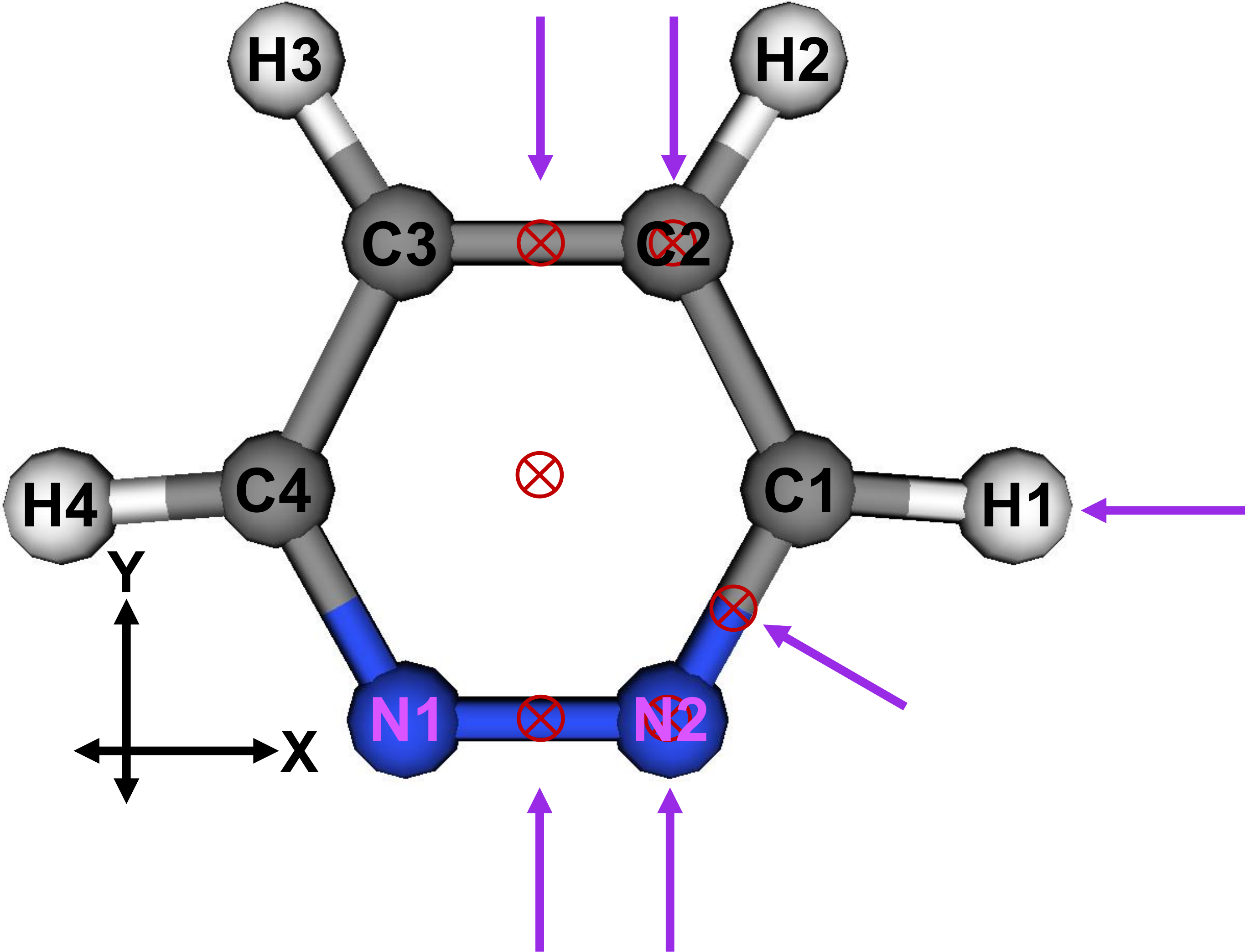}
    \caption{Molecular structure and atom labels for pyridazine (C\textsubscript{4}H\textsubscript{4}N\textsubscript{2}), oriented in the $xy$-plane. The six red ``$\otimes$'' symbols indicate the impact points where the ion approaches orthogonally to the molecular plane. The purple arrows along the sides of the molecule denote impact points for which the projectile travels within the molecular plane.}
    \label{fig:ip-diagram}
\end{figure}

In the following section, we present results from a series of TDDFT simulations in which energetic C$^{5+}$ ions impinge on pyridazine at various impact points (IPs) and orientations. The simulations were performed at zero temperature to remove initial thermal motion of the molecule's atoms as a variable. This system was motivated by the  recent experimental study by Yuan \textit{et al.}~\cite{PhysRevLett.133.193002}. The selected IPs are illustrated in Fig.~\ref{fig:ip-diagram}. Six IPs were chosen to represent projectile trajectories orthogonal to the molecular plane, sampling key regions of the molecule, including individual atoms (C2 and N2), bond centers (C2\textendash C3, C1\textendash N2, and N1\textendash N2), and the molecular center. These points were selected to balance comprehensive coverage of the molecule with computational feasibility. A similar set of IPs were used for simulations in which the projectile moves within the molecular plane. This selective sampling of IPs and projectile orientations aims to approximate experimental conditions, where the site of impact and incident angle are random, while focusing on the most physically relevant scenarios that capture the extrema of molecular responses.

For all simulations, the pyridazine molecule is oriented in the $xy$-plane. In the orthogonal-incidence simulations, the projectile is initially positioned 17.5~\AA\ above the molecular plane, aligned in $x$ and $y$ with the target IP (see Fig.~\ref{fig:ip-diagram}). For the in-plane trajectories, the initial positions of the projectiles are located at least 20~\AA\ from the molecular center, with the incident directions indicated by the heads of the purple arrows in Fig.~\ref{fig:ip-diagram} (tails indicate the initial positions, not drawn to scale). These configurations allow for a controlled study of both orthogonal and planar impacts while capturing realistic variations in projectile approach relevant to experimental setups.

\section{Results}

\begin{figure*}[ht!]
    \centering

    \begin{subfigure}[b]{0.45\textwidth}
        \centering
        \includegraphics[width=\textwidth]{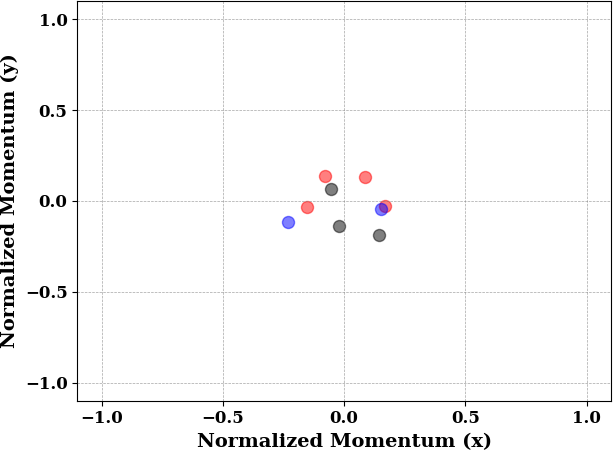}
        \caption{C2}
        \label{fig:pyridazine-c}
    \end{subfigure}
    \hfill
    \begin{subfigure}[b]{0.45\textwidth}
        \centering
        \includegraphics[width=\textwidth]{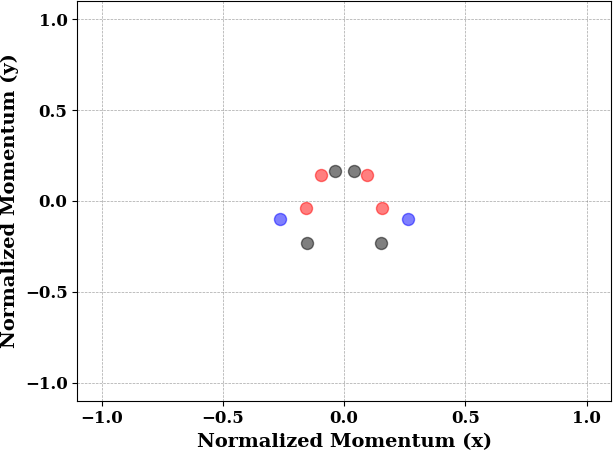}
        \caption{C2\textendash C3}
        \label{fig:pyridazine-cc}
    \end{subfigure}

    \vspace{0.3cm}
    \begin{subfigure}[b]{0.45\textwidth}
        \centering
        \includegraphics[width=\textwidth]{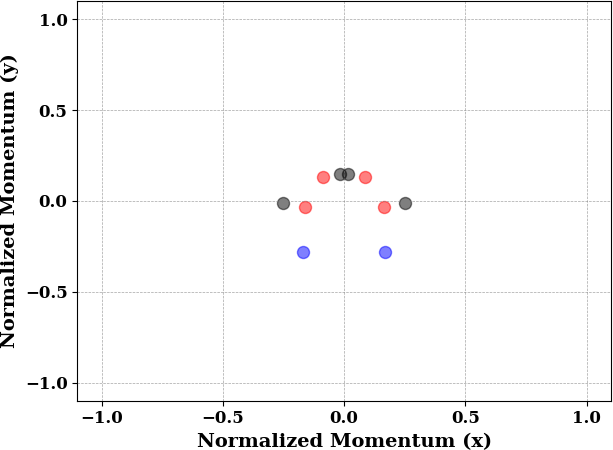}
        \caption{Center}
        \label{fig:pyridazine-center}
    \end{subfigure}
    \hfill
    \begin{subfigure}[b]{0.45\textwidth}
        \centering
        \includegraphics[width=\textwidth]{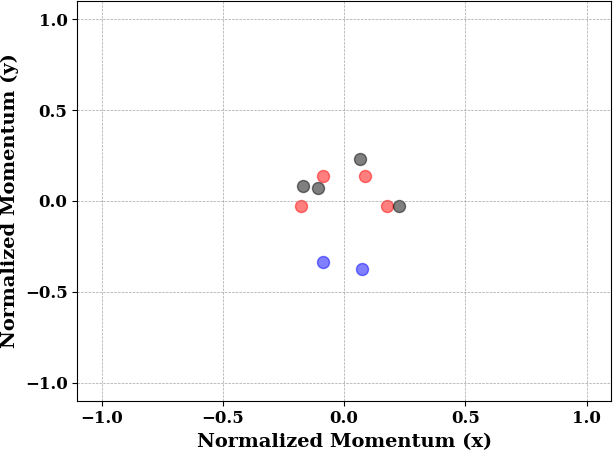}
        \caption{C1\textendash N2}
        \label{fig:pyridazine-cn}
    \end{subfigure}

    \vspace{0.3cm}
    \begin{subfigure}[b]{0.45\textwidth}
        \centering
        \includegraphics[width=\textwidth]{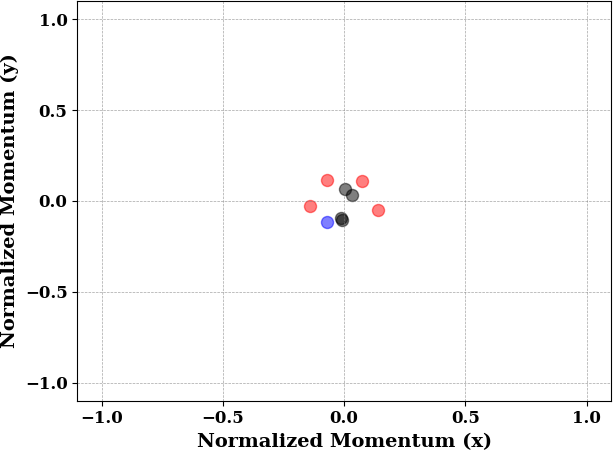}
        \caption{N2}
        \label{fig:pyridazine-n}
    \end{subfigure}
    \hfill
    \begin{subfigure}[b]{0.45\textwidth}
        \centering
        \includegraphics[width=\textwidth]{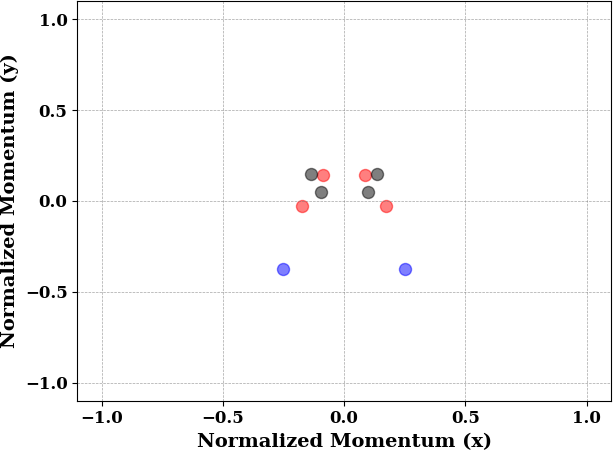}
        \caption{N1\textendash N2}
        \label{fig:pyridazine-nn}
    \end{subfigure}

    \caption{Newton plots showing the normalized $x$- and $y$-components of the ion momenta for pyridazine fragments. Each panel corresponds to a different IP where the projectile strikes the molecule orthogonally to its plane. The subcaption of each panel indicates the corresponding IP. Carbon, hydrogen, and nitrogen atoms are colored gray, red, and blue, respectively.}
    \label{fig:orthogonal_newton_plots}
\end{figure*}

The Newton plots generated from each simulation at each orthogonal impact point (IP) are shown in Fig.~\ref{fig:orthogonal_newton_plots}. Overall, the plots qualitatively resemble the structure of pyridazine, with the momenta of the H, C, and N atoms appearing in similar relative positions. Distinctive features, such as the trapezoidal arrangement formed by the hydrogen atom momenta, are preserved. This agreement is expected: after ionization, the positively charged fragments repel one another, and the resulting momentum distributions reflect the molecular geometry.

The Newton plots exhibit the least distortion when the IP lies on the $y$-axis at $x=0$, corresponding to configurations that preserve the mirror symmetry of the pyridazine molecule (see Fig.~\ref{fig:pyridazine-cc}, \ref{fig:pyridazine-center}, and \ref{fig:pyridazine-nn}). In these cases, the symmetry of the impact conditions aligns with the intrinsic symmetry of the molecule, leading to more faithful reconstruction.

In contrast, the most distorted momentum distributions arise when the projectile ion directly strikes an atom, as seen in Fig.~\ref{fig:pyridazine-c} and Fig.~\ref{fig:pyridazine-n}. Direct atomic impact leads to severe geometric disruption: the projectile transfers a substantial fraction of its kinetic energy to the target atom, ejecting it from the molecular framework with extremely high kinetic energy. Because this energy is much larger than that acquired by the other atoms, the struck atom often leaves the normalized plotting range of the Newton plots in Fig.~\ref{fig:orthogonal_newton_plots}. Consequently, simulations involving direct hits frequently show the impacted atom missing from the reconstructed momentum map.

This behavior is physically accurate: a Newton plot reflects the final momentum vectors of the fragments, and an atom that is violently ejected will not contribute meaningfully to the reconstruction of the molecular structure. This result implies an important limitation for ion-induced CEI under experimental conditions: if the projectile directly impacts (or reaches close enough to) a single atom, the reconstructed molecular geometry may fail to resemble the original structure due to excessive energy transfer to the struck atom.

\begin{figure*}
\centering
\includegraphics[width=\textwidth]{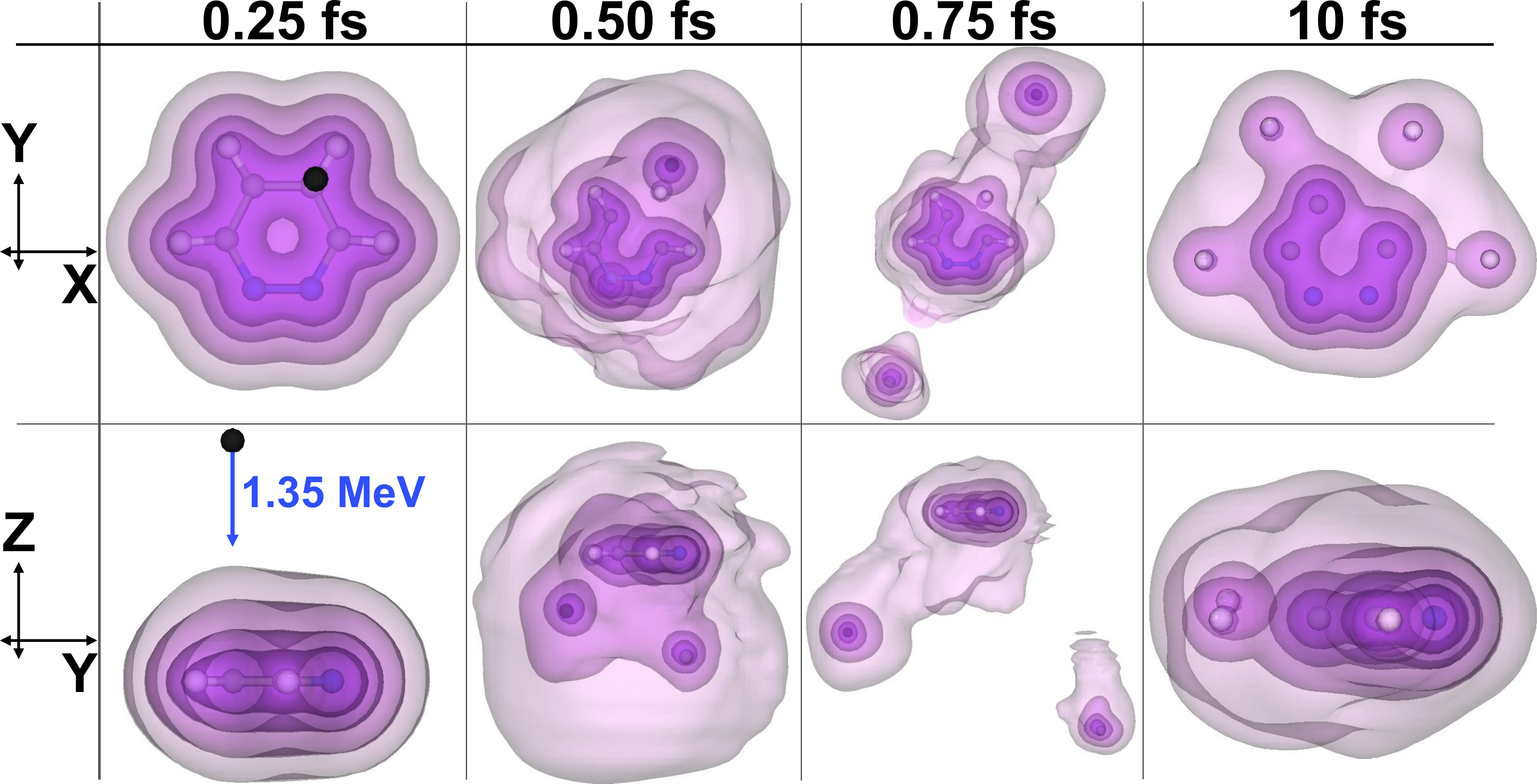}
\caption{Snapshots of the interaction between the C\(^{5+}\) ion (black) and pyridazine for the case where the projectile strikes the C2 atom orthogonal to the molecular plane. Two viewing perspectives are shown: the $xy$-plane (top row) and the $yz$-plane (bottom row). Columns correspond to increasing simulation time. Electron density isosurfaces at values of 0.5, 0.1, 0.01, and 0.001 are rendered in purple.}
\label{fig:c-orthogonal-snapshot}
\end{figure*}
This process of the ion ejecting a target atom from the molecular structure is illustrated in the snapshot diagram corresponding to the C2 IP, with the projectile ion incident orthogonal to the molecular plane (Fig.~\ref{fig:c-orthogonal-snapshot}; see Fig.~\ref{fig:ip-diagram} for the definition of this IP on the pyridazine molecule). As shown in the sequence, the ion approaches the molecule at 0.25~fs with an initial kinetic energy of 1.35~MeV. By 0.5~fs, the ion has transferred a significant portion of its kinetic energy to the C atom at the impact point, dislodging it from the molecular structure while the ion continues forward at a high velocity. By 0.75~fs, both the rapidly moving C atom and the ion have exited the simulation frame, along with the surrounding electron density. This leaves the remaining atoms, now strongly positively charged, to repel one another and undergo Coulomb explosion by 10~fs. 

As shown in Fig.~\ref{fig:c-orthogonal-snapshot}, the interaction between the ion and the nearby N and C atoms not only ejects the C atom but also induces polarization of the electron density as it is drawn toward the projectile. This redistribution of electron density alters the local electrostatic environment and further distorts the resulting momentum images (Fig.~\ref{fig:pyridazine-c}). These observations highlight a key limitation of CEI using highly charged ions: the projectile may interact strongly with the atoms in the molecule, either by imparting excessive kinetic energy to specific atoms or by polarizing the electron cloud, thereby distorting the measured momentum distributions. Experimentally, such effects could manifest as additional noise, increased spread in the momentum images, or greater variability across repeated measurements.

In general, because the projectile interacts differently depending on its point of approach, the most accurate reconstruction of the molecular structure in the Newton plots occurs when the ion does not directly strike an atom but instead passes through a region where it can capture electrons without imparting large forces on specific nuclei. For example, in Fig.~\ref{fig:orthogonal_newton_plots}, the molecular geometry is most faithfully represented for the central IP (Fig.~\ref{fig:pyridazine-center}), where all atoms appear in the momentum map, the distribution is approximately symmetric about the $y$-axis, and the relative positions of key structural features are preserved—such as the four-carbon trapezoidal arrangement with the two nitrogen atoms positioned below it, and the four hydrogen atoms maintaining a similar trapezoidal motif. 

Other Newton plots, however, show clear distortions when specific atoms are targeted. When the C2 atom is struck (Figs.~\ref{fig:pyridazine-c} and \ref{fig:c-orthogonal-snapshot}) or when the N2 atom is impacted (Fig.~\ref{fig:pyridazine-n}), the reconstructed structure contains missing atoms or severely displaced momentum vectors. In cases where bonds are targeted—such as the C2--C3, N1--N2, or C1--N2 bonds—the molecular structure is generally preserved, with no missing atoms. However, the more detailed structural patterns, such as the characteristic trapezoidal arrangement of the carbon atoms and the placement of the nitrogen atoms, are not always captured with full fidelity in these momentum images.

\begin{figure*}[ht!]
    \centering

    \begin{subfigure}[b]{0.45\textwidth}
        \centering
        \includegraphics[width=\textwidth]{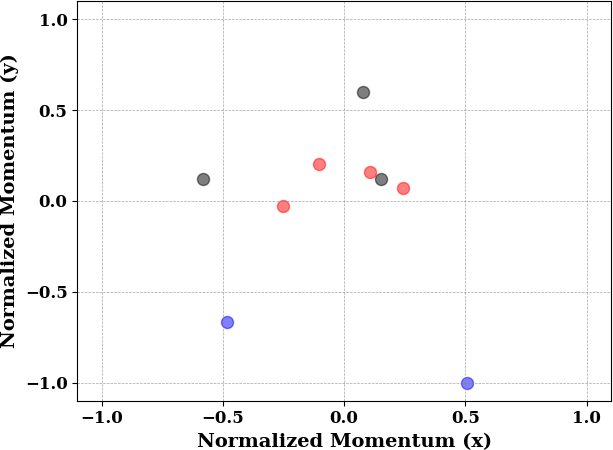}
        \caption{C2 (XY)}
        \label{fig:pyridazine-c-xy}
    \end{subfigure}
    \hfill
    \begin{subfigure}[b]{0.45\textwidth}
        \centering
        \includegraphics[width=\textwidth]{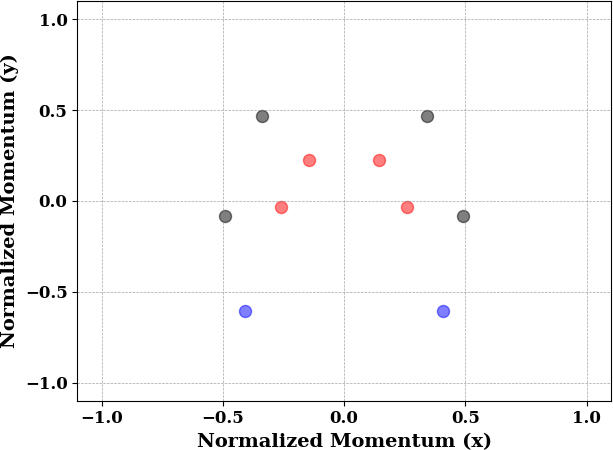}
        \caption{C2\textendash C3 (XY)}
        \label{fig:pyridazine-cc-xy}
    \end{subfigure}

    \vspace{0.3cm}
    \begin{subfigure}[b]{0.45\textwidth}
        \centering
        \includegraphics[width=\textwidth]{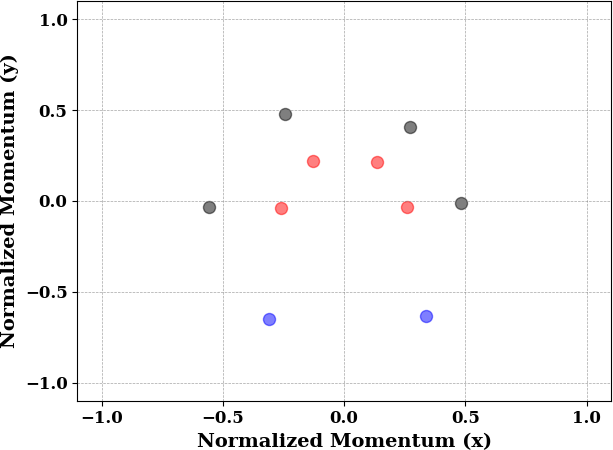}
        \caption{C1\textendash N2 (XY)}
        \label{fig:pyridazine-cn-xy}
    \end{subfigure}
    \hfill
    \begin{subfigure}[b]{0.45\textwidth}
        \centering
        \includegraphics[width=\textwidth]{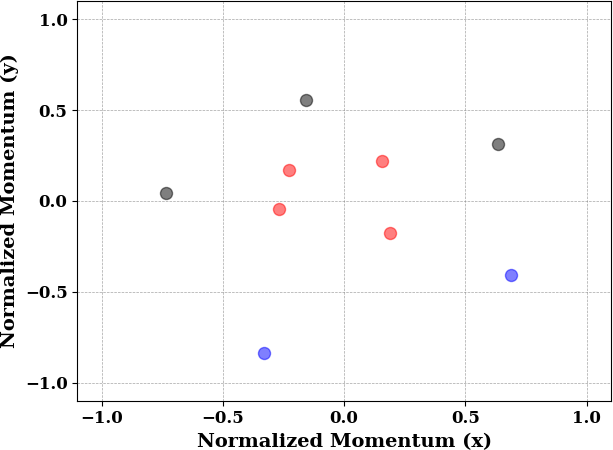}
        \caption{H1 (XY)}
        \label{fig:pyridazine-h-xy}
    \end{subfigure}

    \vspace{0.3cm}
    \begin{subfigure}[b]{0.45\textwidth}
        \centering
        \includegraphics[width=\textwidth]{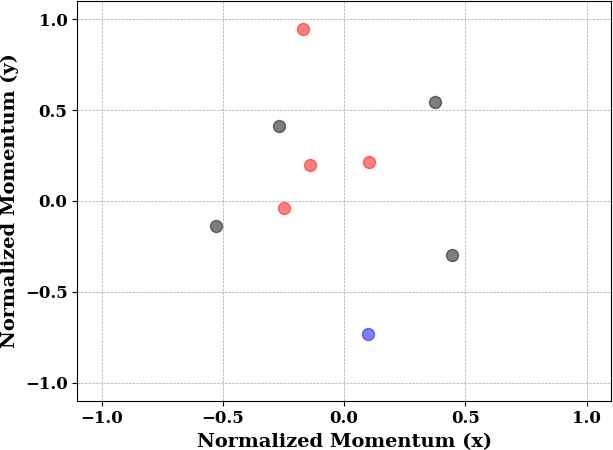}
        \caption{N2 (XY)}
        \label{fig:pyridazine-n-xy}
    \end{subfigure}
    \hfill
    \begin{subfigure}[b]{0.45\textwidth}
        \centering
        \includegraphics[width=\textwidth]{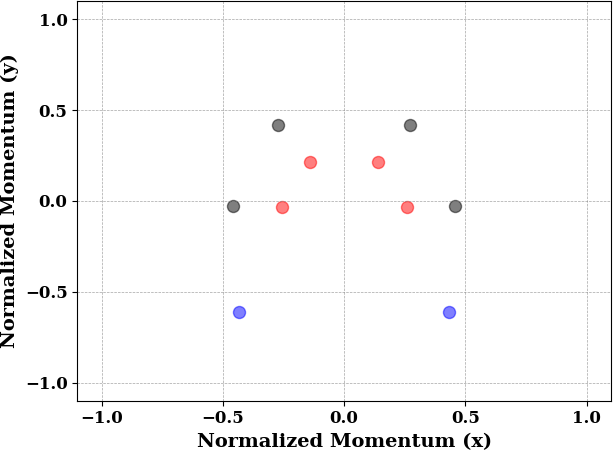}
        \caption{N1\textendash N2 (XY)}
        \label{fig:pyridazine-nn-xy}
    \end{subfigure}

    \caption{Newton plots showing the normalized $x$- and $y$-components of the ion momenta for pyridazine fragments. Each panel corresponds to a different IP where the projectile travels within the molecular XY plane, striking the molecule in-plane. The subcaption for each panel specifies the corresponding IP.}
    \label{fig:xy-plane-newton-plots}
\end{figure*}

The Newton plots generated for simulations in which the projectile begins and travels within the molecular plane (the $xy$-plane in our calculations) are shown in Fig.~\ref{fig:xy-plane-newton-plots} and reveal trends similar to those observed in the orthogonal-incidence case. In these in-plane trajectories, the IPs where the projectile travels along and targets the $y$-axis—such as when the C2--C3 or N1--N2 bonds are struck (Figs.~\ref{fig:pyridazine-cc-xy} and \ref{fig:pyridazine-nn-xy}, respectively)—yield the most accurate reconstruction of the molecular structure. This is expected: the system possesses mirror symmetry about the $y$-axis, and because the projectile’s initial position and direction of motion lie along this axis, that symmetry is preserved throughout the interaction.

In these symmetric cases, the projectile exerts comparable forces on atoms located on either side of the molecule, preventing asymmetric distortion. Additionally, no significant electron-density polarization occurs along the $x$- or $z$-directions; any redistribution of electron density is constrained primarily to the $y$-axis. This suppression of lateral perturbations allows the momentum distributions to more faithfully resemble the underlying molecular geometry.

\begin{figure*}
\centering
\includegraphics[width=\textwidth]{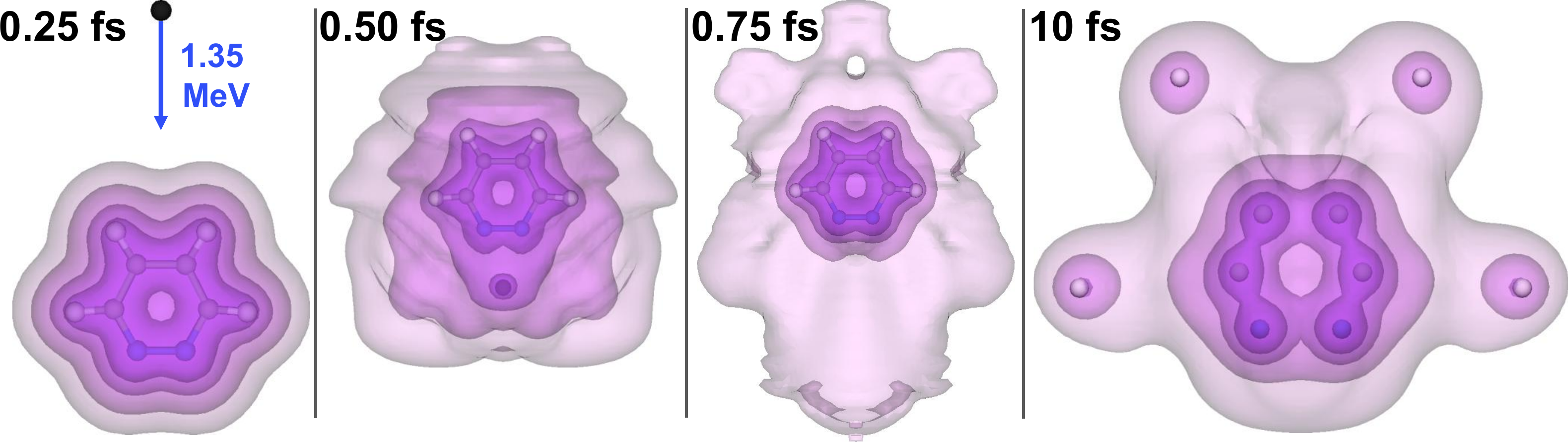}
\caption{Snapshots of the interaction between the C\(^{5+}\) ion and pyridazine when the projectile strikes the C2--C3 bond within the molecular XY plane. Electron density isosurfaces at values of 0.5, 0.1, 0.01, and 0.001 are shown in purple.}
\label{fig:cc-xy-plane-snapshot}
\end{figure*}

The snapshots of the Coulomb explosion for the trajectory in which the projectile targets the C2--C3 bond are shown in Fig.~\ref{fig:cc-xy-plane-snapshot}. As depicted, the ion approaches the molecule and begins attracting nearby electron density by approximately 0.25~fs. By 0.5~fs, the projectile has passed through the molecular plane, and electron-density integration reveals that it has captured approximately 3.11 electrons from the molecule. By 0.75~fs, the ion has reached the CAP region and exited the simulation frame while preserving symmetry about the $y$-axis, with no significant electron-density polarization along the $x$- or $z$-directions.

By 10~fs (as shown in Fig.~\ref{fig:pyridazine-cc-xy}), the remaining molecular framework undergoes Coulomb explosion, maintaining near-perfect symmetry about the $y$-axis; the positively charged atoms repel radially outward, as reflected in the corresponding Newton plot (Fig.~\ref{fig:pyridazine-cc-xy}). At this stage, electron-density analysis indicates that a total of 7.57 electrons have been removed from the pyridazine molecule. Because the projectile itself captures only about 3.11 electrons, the results imply that ionization proceeds through a combination of electron capture by the projectile and additional electron ejection from the molecular system.

\begin{figure*}
\centering
\includegraphics[width=\textwidth]{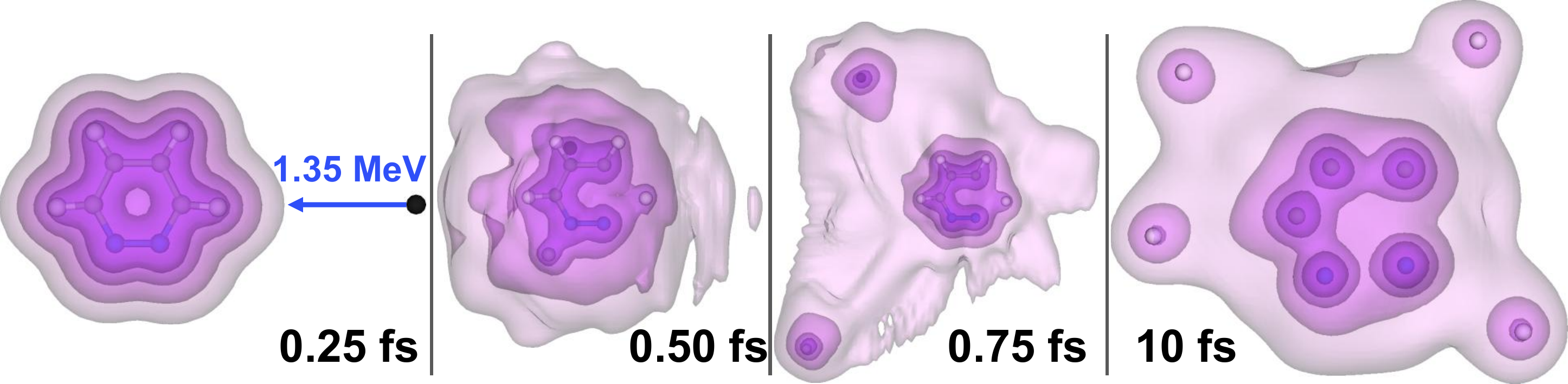}
\caption{Snapshots of the interaction between the C\(^{5+}\) ion and pyridazine when the projectile is directed at the H1 atom within the molecular XY plane. Electron density isosurfaces at values of 0.5, 0.1, 0.01, and 0.001 are displayed in purple.}
\label{fig:h-xy-plane-snapshot}
\end{figure*}

Analyzing cases in which the projectile is directed toward an individual atom from within the molecular plane also provides valuable insight. For example, Fig.~\ref{fig:h-xy-plane-snapshot} shows the snapshots for the trajectory aimed at the H1 atom with the projectile propagating in the $xy$-plane. At 0.25~fs, the ion approaches the molecule, and by 0.5~fs it bypasses the target H1 atom and instead knocks the C1 atom outward toward the lower-left N1 atom. Simultaneously, the ion is deflected upward near the upper-left H3 atom. By 0.75~fs, both the fast-moving C$^{5+}$ ion and the displaced C1 atom have exited the frame. The remaining positively charged atoms repel one another, and by 10~fs the resulting structure resembles that shown in the corresponding Newton plot in Fig.~\ref{fig:pyridazine-h-xy}.

The Newton plots for projectiles traveling within the molecular plane (Fig.~\ref{fig:xy-plane-newton-plots}) reveal trends consistent with those observed in the orthogonal-incidence case. When bonds or non-atomic centers are targeted, the resulting molecular structure is generally clearer and more accurately reconstructed. However, when individual atoms are directly struck, the Newton plots become noticeably more distorted. For instance, when the right-side H1 atom is hit, that hydrogen is displaced significantly downward in the momentum plot, exhibiting a much larger negative $y$-momentum than in other cases, and the adjacent C1 atom is entirely ejected from the system.

A comparison between the momentum images produced by orthogonal projectiles (Fig.~\ref{fig:orthogonal_newton_plots}) and in-plane projectiles (Fig.~\ref{fig:xy-plane-newton-plots}) shows that in-plane trajectories generate a wider spread in the atomic momenta. This effect arises from increased ionization: the average ionization at 10~fs is 4.78 electrons for orthogonal trajectories, but 6.95 electrons for planar trajectories. The additional removal of roughly 2.17 electrons leaves the molecular fragments more positively charged, increasing the Coulomb repulsion and leading to higher fragment velocities.

This enhanced ionization for in-plane trajectories is attributed to the longer interaction time between the projectile and the molecular electronic density. When the ion travels within the molecular plane, it traverses a larger cross-sectional region of the electron cloud, allowing it to spend more time near multiple atomic sites. This enables the ion to capture additional electrons localized around these atoms and simultaneously repel nearby positively charged nuclei more strongly, thereby amplifying both ionization and fragment acceleration.

\begin{figure}[ht!]
    \centering
    \includegraphics[width=0.95\columnwidth]{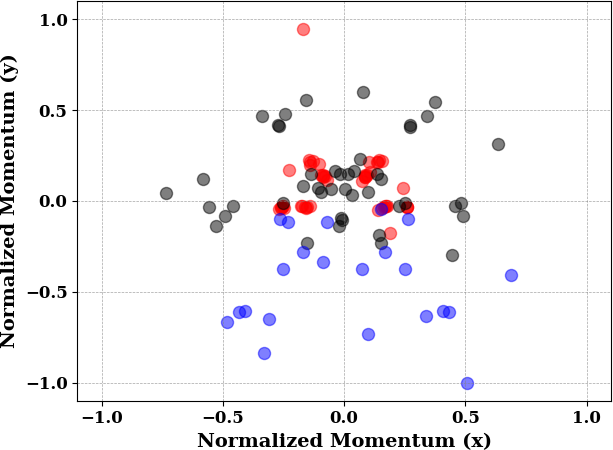}
    \caption{Full Newton plot combining the momentum images from all 12 simulations across all IPs, showing the overall distribution of atomic momenta for pyridazine fragments.}
    \label{fig:full-newton-plot}
\end{figure}

Combining all of the individual Newton plots for each IP (Figs.~\ref{fig:orthogonal_newton_plots} and \ref{fig:xy-plane-newton-plots}) into a single plot is shown in Fig.~\ref{fig:full-newton-plot}. This combined representation mirrors how a typical Newton plot is constructed experimentally: by overlaying a large number of data points on one graph so that, given sufficient statistics, the underlying molecular structure emerges.

The full Newton plot in Fig.~\ref{fig:full-newton-plot} illustrates how widely distributed the atomic momenta are across all simulations. The characteristic trapezoidal pattern typically formed by the H atoms (in red) is still visible. The C atoms (in gray), however, appear to form both an inner and an outer semicircle, reflecting a noticeable gap between two momentum populations. This separation arises from combining the in-plane and out-of-plane projectile simulations\textemdash an effect also evident in the N atoms (in blue). The atoms with the largest momenta in these distributions originate from the planar trajectories, consistent with the earlier analysis.

In experimental CEI, because a large ensemble of ionization events is accumulated over many stochastic molecular orientations and IPs, the resulting Newton plots tend to reveal the molecular geometry more clearly, with the average structure appearing as the densest region of the distribution. In contrast, the present study focuses on extreme, controlled cases\textemdash strictly orthogonal versus planar projectile orientations, and precisely defined IPs. As a result, the combined Newton plot presented here represents something closer to the \emph{maximum} possible spread in the momentum distribution that can arise from varying these IP and orientation parameters.


\begin{table*}[t]
\centering
\renewcommand{\arraystretch}{1.25}
\setcellgapes{3pt}\makegapedcells
\begin{tabular*}{\textwidth}{@{\extracolsep{\fill}}l l c c c c c c c c c c c}
\toprule
\textbf{IP} &
\textbf{Geom.} &
\textbf{C1} & \textbf{C2} & \textbf{C3} & \textbf{C4} & \textbf{H1} & \textbf{H2} & \textbf{H3} & \textbf{H4} & \textbf{N1} & \textbf{N2} & \textbf{Total} \\
\midrule
\arrayrulecolor[gray]{0.75}
\textbf{C2} & Orthogonal & 0.49 & \textemdash & 0.52 & 0.40 & 0.78 & 0.71 & 0.65 & 0.65 & 0.30 & 0.16 & 4.66 \\ \midrule
\textbf{C2} & Planar & 0.72 & \textemdash & 0.85 & 0.79 & 0.70 & 0.59 & 0.58 & 0.70 & 0.56 & 0.76 & 6.25 \\ \midrule
\textbf{N2} & Orthogonal & 0.44 & 0.44 & 0.29 & 0.30 & 0.63 & 0.50 & 0.53 & 0.69 & 0.36 & \textemdash & 4.18 \\ \midrule
\textbf{N2} & Planar & 0.88 & 0.85 & 0.72 & 0.99 & \textemdash & 0.58 & 0.54 & 0.66 & 0.65 & \textemdash & 5.88 \\ \midrule
\textbf{C2\textendash C3} & Orthogonal & 0.36 & 0.40 & 0.40 & 0.36 & 0.64 & 0.66 & 0.66 & 0.64 & 0.40 & 0.40 & 4.94 \\ \midrule
\textbf{C2\textendash C3} & Planar & 0.78 & 0.86 & 0.86 & 0.78 & 0.74 & 0.73 & 0.73 & 0.74 & 0.67 & 0.67 & 7.57 \\ \midrule
\textbf{C1\textendash N2} & Orthogonal & 0.58 & 0.25 & 0.41 & 0.49 & 0.73 & 0.66 & 0.54 & 0.72 & 0.25 & 0.36 & 4.98 \\ \midrule
\textbf{C1\textendash N2} & Planar & 0.87 & 0.88 & 0.39 & 0.75 & 0.79 & 0.69 & 0.70 & 0.81 & 0.57 & 0.89 & 7.34 \\ \midrule
\textbf{N1\textendash N2} & Orthogonal & 0.41 & 0.36 & 0.36 & 0.41 & 0.66 & 0.62 & 0.62 & 0.66 & 0.41 & 0.41 & 4.94 \\ \midrule
\textbf{N1\textendash N2} & Planar & 0.53 & 0.93 & 0.93 & 0.53 & 0.77 & 0.73 & 0.73 & 0.77 & 0.72 & 0.72 & 7.38 \\ \midrule
\textbf{Center} & Orthogonal & 0.38 & 0.39 & 0.39 & 0.38 & 0.73 & 0.68 & 0.68 & 0.73 & 0.31 & 0.31 & 4.96 \\ \midrule
\textbf{H1} & Planar & \textemdash & 1.16 & 0.67 & 0.64 & 0.68 & 0.81 & 0.72 & 0.69 & 1.00 & 0.93 & 7.30 \\ \midrule
\textbf{Mean} & Orthogonal & 0.45 & 0.37 & 0.39 & 0.39 & 0.70 & 0.64 & 0.61 & 0.68 & 0.34 & 0.33 & 4.78 \\ \midrule
\textbf{Mean} & Planar & 0.76 & 0.94 & 0.74 & 0.75 & 0.74 & 0.69 & 0.67 & 0.73 & 0.70 & 0.79 & 6.95 \\ \midrule
\textbf{Mean} & All & 0.59 & 0.65 & 0.57 & 0.57 & 0.71 & 0.66 & 0.64 & 0.71 & 0.52 & 0.56 & 5.87 \\
\arrayrulecolor{black}
\bottomrule
\end{tabular*}
\caption{Ionization of each atom for different IPs and projectile orientations (orthogonal to the molecular plane or within the plane). The symbol ``\textemdash'' indicates that the atom reached the complex absorbing potential (CAP) by 10~fs and therefore carried no electron density. The last three rows report the mean ionization for each geometry: the orthogonal mean is calculated over all orthogonal IPs, the planar mean over all planar IPs, and the overall mean across all simulations. Atoms that exited the CAP are excluded from the mean calculations.
}
\label{tab:ionization}
\end{table*}

For each IP and orientation, the ionization (number of valence electrons lost) from each atom in C\textsubscript{4}H\textsubscript{4}N\textsubscript{2} remaining in the simulation box at 10~fs is presented in Table~\ref{tab:ionization}. The final column gives the total electron loss for each trajectory, while the last three rows provide mean values across the orthogonal, planar, and combined sets of simulations. A clear and systematic trend emerges: orthogonal impacts consistently generate less ionization than planar impacts\textemdash 4.78 versus 6.95 electrons on average. This difference of 2.17 electrons aligns with the behavior observed in the Newton plots, where planar geometries exhibit the largest fragment momenta.

The orientation dependence is also evident at the single-atom level. Yuan \textit{et al.}~\cite{PhysRevLett.133.193002} detected H$^{+}$, C$^{2+}$, C$^{+}$, and N$^{+}$ fragments in quadruple coincidence and inferred a minimum total charge of $+11$, corresponding to assigning one carbon atom a charge of $+2$ and all remaining atoms $+1$. By contrast, our TDDFT calculations, which track fractional electron loss rather than imposing integer charge states, yield an average total ionization of only 5.78 electrons across all IPs and orientations.

This discrepancy has several physical origins. For instance, reaction-microscope measurements are intrinsically sensitive only to charged fragments: the extraction fields guide ions and electrons to the detector, while any neutral fragments continue ballistically and remain undetected~\cite{PhysRevLett.133.193002}. Consequently, the experimentally reconstructed charge corresponds to the \emph{minimum} integer charge state compatible with the observed set of ionic fragments. Any neutrals produced during the explosion leave no signature on the detector, meaning the true degree of ionization in the experiment may be lower than the inferred value, and a direct one-to-one correspondence with TDDFT electron counts is neither expected nor appropriate.

Additional experimental factors further increase the effective ionization. In the measurement of Ref.~\cite{PhysRevLett.133.193002}, each molecule interacts not with a single isolated projectile, but with a continuous beam of C$^{5+}$ ions. Long-range Coulomb fields from neighboring projectiles and from the beam’s collective space charge can strip additional electrons or suppress electron recapture after the primary collision. Processes such as post-collision ionization~\cite{PhysRevA.106.033114}, shake-off~\cite{annurev:/content/journals/10.1146/annurev.ns.24.120174.001233, https://doi.org/10.1110/ps.ps.26201}, and correlated multielectron dynamics~\cite{BERAKDAR200391}\textemdash occurring tens to hundreds of femtoseconds after the initial impact\textemdash also contribute to the final charge state but lie well outside the 10~fs window accessible to TDDFT.

By contrast, our simulations model a \emph{single} projectile–molecule interaction within a finite simulation domain. Atoms that travel into the complex absorbing potential (CAP) are removed from the calculation and cannot undergo secondary stripping or late-time interactions from additional ion projectiles in the beam. As a result, TDDFT naturally yields lower total electron removal than the experimentally inferred integer charges reconstructed from reaction-microscope data.

Taken together, Table~\ref{tab:ionization} provides a quantitative map of how ionization varies with projectile orientation, impact point, and atomic site. These trends reinforce the structural behavior observed in the Newton plots (Figs.~\ref{fig:orthogonal_newton_plots}, \ref{fig:xy-plane-newton-plots}, \ref{fig:full-newton-plot}) and in the snapshot diagrams (Figs.~\ref{fig:c-orthogonal-snapshot}, \ref{fig:cc-xy-plane-snapshot}, \ref{fig:h-xy-plane-snapshot}). They also clarify why TDDFT underestimates the integer ionization inferred from experiment: the simulations capture prompt electron removal in a single-collision geometry, whereas the experiment measures the minimal integer charge of the surviving ionic fragments after additional late-time and beam-induced processes.

\section{Summary}

In this work, we investigated the Coulomb explosion of pyridazine (C\textsubscript{4}H\textsubscript{4}N\textsubscript{2}) induced by energetic C$^{5+}$ projectiles using time-dependent density-functional theory (TDDFT) combined with Ehrenfest nuclear dynamics. By systematically examining orthogonal and in-plane projectile trajectories across multiple impact points (IPs), we characterized how both projectile orientation and point of incidence govern fragment momenta, electron-density response, and atom-specific ionization.

Newton plots (Figs.~\ref{fig:orthogonal_newton_plots}, \ref{fig:xy-plane-newton-plots}, and \ref{fig:full-newton-plot}) showed that trajectories avoiding direct atomic collisions produce the most faithful structural reconstructions, preserving the relative arrangement of H, C, and N atoms and maintaining recognizable molecular motifs. In contrast, direct atomic impacts lead to strong distortions: the struck atoms acquire disproportionately large momenta and degrade the quality of structural recovery. Planar trajectories, which pass through a larger portion of the molecular electron cloud, generate the largest overall momentum spreads due to their inherently higher ionization.

Electron-density snapshots (Figs.~\ref{fig:c-orthogonal-snapshot}, \ref{fig:cc-xy-plane-snapshot}, and \ref{fig:h-xy-plane-snapshot}) further elucidate these dynamics. The traversing projectile polarizes the surrounding electron density, redistributes charge toward nearby atoms, and initiates local bond weakening or rupture. In-plane trajectories remove substantially more electrons than orthogonal ones, increasing Coulomb repulsion and enhancing fragment velocities, while orthogonal trajectories generate lower ionization and correspondingly smaller momentum dispersions.

Quantitative ionization analysis (Table~\ref{tab:ionization}) reinforces these trends. Orthogonal impacts yield an average of 4.78 electrons removed, compared to 6.95 for planar impacts, with clear orientation-dependent variations at the level of individual atoms. The combined Newton plot (Fig.~\ref{fig:full-newton-plot}) illustrates the overall momentum envelope produced across all simulations while preserving the recognizable pyridazine geometry.

Overall, this study demonstrates that projectile orientation and impact geometry play essential roles in shaping fragment momenta, electron-density dynamics, and atom-resolved ionization. These factors directly influence the fidelity of structural reconstruction in Coulomb explosion imaging (CEI) and help explain sources of variability, noise, and charge-state discrepancies observed in experimental ion-induced CEI datasets. At the same time, the results highlight key limitations of this emerging approach: its strong dependence on the molecule’s instantaneous orientation, its sensitivity to variations in impact position, and its propensity for localized regions of the target to be disproportionately disrupted. Together, these effects broaden the resulting momentum distributions and introduce 
additional noise in the reconstructed structural images.  

Our study employed the Ehrenfest mean-field method, in which atomic
nuclei move according to an averaged potential energy surface obtained
by weighting contributions from all electronic states by their
populations. Alternative methodologies exist: trajectory surface
hopping (TSH) represents nuclear wavepacket behavior through
collections of independent trajectories, each confined to a single
adiabatic surface, while multiple spawning (MS) builds the nuclear
wavefunction from Gaussian basis functions that move along classical
paths. Although MS captures quantum mechanical nuclear motion, all
three methods depend on a common prerequisite—calculating electronic
structure properties (energy gradients, nonadiabatic coupling terms)
at classical nuclear configurations during time evolution. A
comprehensive analysis of the strengths and limitations of each
technique appears in Reference \cite{doi:10.1021/acs.chemrev.7b00577}.
We intend to integrate TSH into our workflow for comparative
assessment against Ehrenfest dynamics.
Given these considerations, the present approach offers a
computationally efficient and scalable strategy for obtaining
qualitative understanding of how reactions proceed and fragment,
though quantitatively accurate result should not be anticipated.

Several avenues of investigation could further extend this study. First, simulations incorporating \emph{multiple} ionic projectiles or long-range fields would better approximate the environment of an actual ion beam and capture beam-induced stripping or suppression of electron recapture. Second, sampling randomized molecular orientations and impact points would generate statistical momentum and ionization distributions directly comparable to experimental CEI events, enabling a more complete assessment of reconstruction fidelity and noise sources. Finally, experimental validation of the predicted atom-specific ionization patterns, momentum distributions, and orientation-dependent effects would provide an important test of the TDDFT framework and strengthen 
the connection between theory and ultrafast CEI measurements.
Beyond aiding experimental interpretation, these calculations have the
potential to drive progress in generative modeling approaches for
molecular-structure retrieval from Coulomb explosion imaging
\cite{li2025,venkatachalam2025}.

\begin{acknowledgments} 
This work has been supported by the National Science Foundation (NSF)
under Grant No. DMR-2217759 and IRES-2245029.

This material is based upon work supported by the National Science Foundation Graduate 
Research Fellowship Program under Grant No. 2140001. Any opinions, 
findings, and conclusions or recommendations expressed in this material are those of the 
author(s) and do not necessarily reflect the views of the National Science Foundation.

This work used ACES at TAMU through allocation PHYS240167 from the Advanced Cyberinfrastructure Coordination Ecosystem: Services \& Support (ACCESS) program, which is supported by National Science 
Foundation grants \#2138259, \#2138286, \#2138307, \#2137603, and \#2138296~\cite{aces}.

\end{acknowledgments}


%

\end{document}